\documentclass[prl,twocolumn,showpacs,floatfix,amsmath,nofootinbib,amssymb,floatfix]{revtex4}
\usepackage{graphicx,color,dcolumn,booktabs,bm,pifont}
\usepackage{longtable,lscape}
\usepackage{txfonts}
\usepackage{overpic}
\usepackage{amssymb}
\usepackage{indentfirst}
\usepackage{feynmf}   
\usepackage{slashed}  
\usepackage{cases}
\usepackage{color}
\usepackage{multirow}
\usepackage{epstopdf}
\usepackage{longtable}
\usepackage{graphicx,color,dcolumn,booktabs,bm}
\usepackage[colorlinks,
            citecolor=blue,
            anchorcolor=red,
            menucolor=red,
            linkcolor=red,
            filecolor=red,
            runcolor=red,
            urlcolor=blue,
            frenchlinks=red]{hyperref}

\graphicspath{{Figures/}} %

\allowdisplaybreaks

\begin{document}

\title{Strong LHCb evidence supporting the existence of the hidden-charm molecular pentaquarks}

\author{Rui Chen$^{1,2,3}$}
\author{Zhi-Feng Sun$^{1,2}$}
\author{Xiang Liu$^{1,2}$}
\email{xiangliu@lzu.edu.cn}
\author{Shi-Lin Zhu$^{3,4,5}$}
\email{zhusl@pku.edu.cn}

\affiliation{
$^1$School of Physical
Science and Technology, Lanzhou University, Lanzhou 730000, China\\
$^2$Research Center for Hadron and CSR Physics, Lanzhou University
and Institute of Modern Physics of CAS, Lanzhou 730000, China\\
$^3$ School of Physics and State Key Laboratory of Nuclear Physics and Technology, Peking University, Beijing 100871, China\\
$^4$Collaborative Innovation Center of Quantum Matter, Beijing 100871, China\\
$^5$Center of High Energy Physics, Peking University, Beijing
100871, China}

\begin{abstract}
On March 26th, 2019, at the Rencontres de Moriond QCD conference,
the LHCb Collaboration reported the observation of three new
pentaquarks, namely $P_c(4312)$, $P_c(4440)$ and $P_c(4457)$, which
are consistent with the loosely bound molecular hidden-charm
pentaquark states composed of an S-wave charmed baryon $\Sigma_c$
and an S-wave anti-charmed meson ($\bar{D}, \bar{D}^*$). In this
work, we present a direct calculation by the one-boson-exchange
(OBE) model and demonstrate explicitly that the $P_c(4312)$,
$P_c(4440)$ and $P_c(4457)$ do correspond to the loosely bound
$\Sigma_c\bar{D}$ with $(I=1/2,J^P=1/2^-)$, $\Sigma_c\bar{D}^*$ with
$(I=1/2,J^P=1/2^-)$ and $\Sigma_c\bar{D}^*$ with
$(I=1/2,J^P=3/2^-)$, respectively. 

\end{abstract}


\maketitle

{\it Introduction}:---Finding the multiquark matter is an extremely
important issue of hadron physics, which is full of challenge and
opportunity not only for theorists but also for experimentalists. In
the past sixteen years, a series of observations of the
charmoniumlike $XYZ$ states and two $P_c$ states have inspired
extensive investigations of the hidden-charm tetraquarks and
pentaquarks, which enlarge our knowledge of the multiquark matter
(see review articles \cite{Chen:2016qju,Liu:2013waa,Guo:2017jvc} for
more details).

Focusing on hidden-charm pentaquark, we must mention the predictions
for the existence of the molecular type pentaquark
\cite{Wu:2010jy,Yang:2011wz,Wang:2011rga,Wu:2012md,Li:2014gra}
before LHCb's observation \cite{Aaij:2015tga}. In 2015, LHCb
announced the observation of two hidden-charm pentaquarks
$P_c(4380)$ and $P_c(4450)$ in the $J/\psi p$ invariant mass
spectrum of $\Lambda_b\to J/\psi K p$ \cite{Aaij:2015tga}. This
observation had stimulated extensive discussions on decoding their
inner structures based on different configurations, which include
the loosely bound molecular baryon-meson state
\cite{Chen:2015loa,Chen:2015moa,Karliner:2015ina} and tightly bound
pentaquark state
\cite{Maiani:2015vwa,Lebed:2015tna,Anisovich:2015cia,Wang:2015ava}. The
experimental data in 2015 was unable to distinguish them.

\begin{figure}[htbp]
\includegraphics[width=200pt]{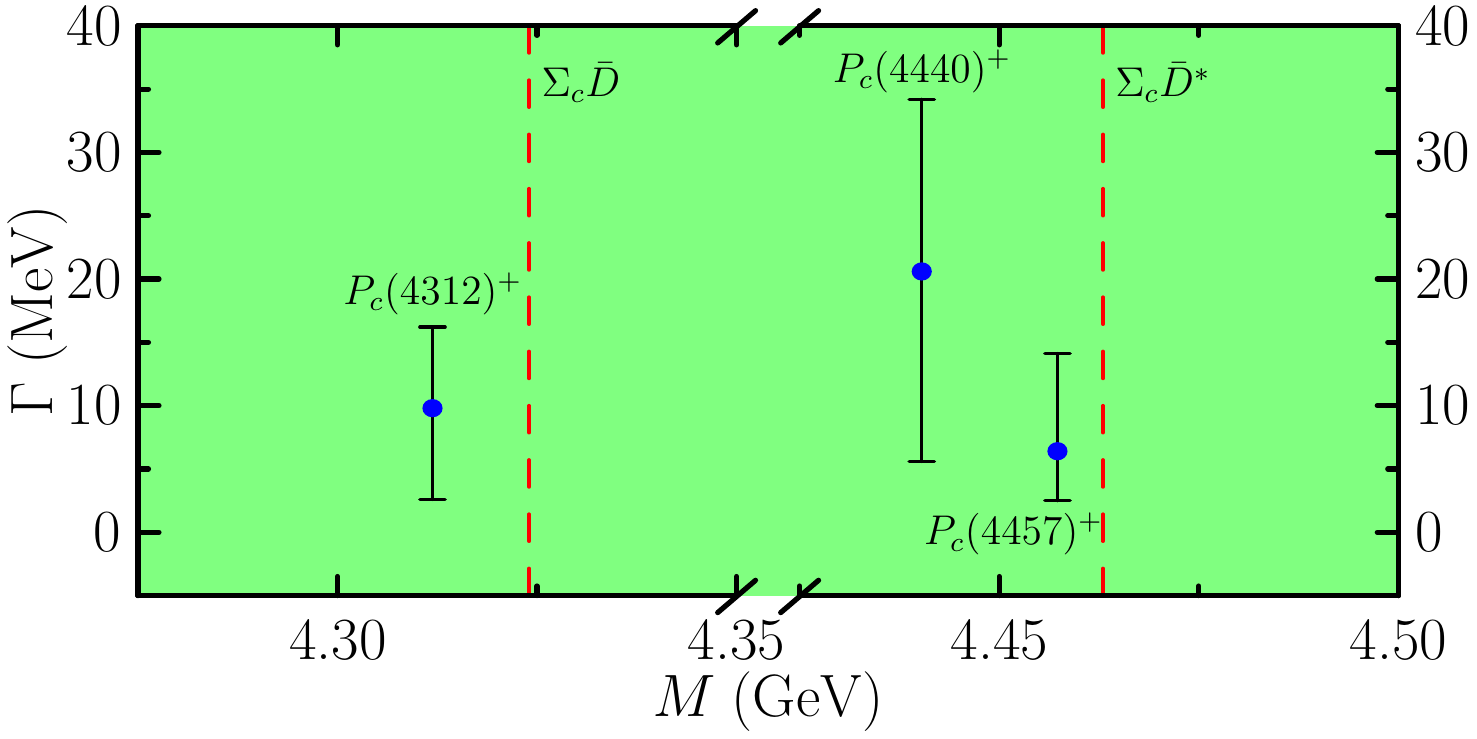}
\caption{The observed three pentaquarks from LHCb and comparison
with the thresholds.}\label{fig11}
\end{figure}

Exploring the novel multiquark matter continues. Very recently, the
LHCb Collaboration once again brought us surprise. On March 26th,
2019, at the Rencontres de Moriond QCD conference, the LHCb
Collaboration reported the observation of three new pentaquarks
\cite{LHCbtalk}. By analyzing the $J/\psi p$ invariant mass
spectrum, a new pentaquark named as the $P_c(4312)$ was discovered
with $7.3\sigma$ significance. And LHCb updated analysis further
indicates that the $P_c(4450)$ reported previously by LHCb
\cite{Aaij:2015tga} contains two narrow subpeaks, $P_c(4440)$ and
$P_c(4457)$, which have 5.4$\sigma$ significance. Their resonance parameters
include  \cite{Aaij:2019vzc} 
\begin{eqnarray*}
 P_c(4312)^+:\quad\left\{
\begin{array}{l}
m=4311.9\pm0.7^{+6.8}_{-0.6}\,{\rm MeV}\\
\Gamma=9.8\pm2.7^{+3.7}_{-4.5}\,{\rm MeV}
\end{array}\right. ,\\
P_c(4440)^+:\quad \left\{
\begin{array}{l}
m=4440.3\pm1.3^{+4.1}_{-4.7}\,{\rm MeV}\\
\Gamma=20.6\pm4.9^{+8.7}_{-10.1}\,{\rm MeV}
\end{array}\right. ,\\
P_c(4457)^+:\quad\left\{
\begin{array}{l}
m=4457.3\pm1.3^{+0.6}_{-4.1}\,{\rm MeV}\\
\Gamma=6.4\pm2.0^{+5.7}_{-1.9}\,{\rm MeV}
\end{array}\right. .
\end{eqnarray*}
Since these three pentaquarks are observed in the $J/\psi p$
invariant mass spectrum, they have definite isospin quantum number
$I=1/2$. We notice that the $P_c(4312)^+$ is just below the
$\Sigma_c^+\bar{D}^0$ threshold, while the masses of the
$P_c(4440)^+$ and $P_c(4457)^+$ are slightly lower than the
$\Sigma_c^+\bar{D}^{*0}$ threshold (see Fig. \ref{fig11}).

By borrowing the experience of studying nuclear force, the OBE model
was developed to investigate these observed charmoniumlike $XYZ$
states
\cite{Liu:2007bf,Liu:2008xz,Liu:2008fh,Liu:2008tn,Liu:2009ei,Close:2009ag}.
Later, in 2011, we applied the OBE model to study the
$\Sigma_c\bar{D}$ and $\Sigma_c\bar{D}^*$ interactions
\cite{Yang:2011wz}. Although the treatment of the $\Sigma_c\bar{D}$
and $\Sigma_c\bar{D}^*$ systems in Ref. \cite{Yang:2011wz} was quite
simple and straightforward, the existence of the $\Sigma_c\bar{D}$
and $\Sigma_c\bar{D}^*$ molecular hidden-charm pentaquark was
predicted clearly, which has become one of the pioneering papers
exploring the molecular hidden-charm pentaquarks.

Until now, the OBE model has been greatly improved compared with the
situation in 2011 (see review \cite{Chen:2016qju}), which stimulates
us to include the coupled-channel effect and S-D wave mixing in the
calculation. Especially, the new observations of three pentaquarks
announced by LHCb makes the study of the hidden-charm molecular
pentaquarks enter upon a new era with higher precision.

In this letter, we will present a precise investigation of the
$\Sigma_c\bar{D}$ and $\Sigma_c\bar{D}^*$ systems, from which the
readers can learn why this baryon-meson configuration of the
hidden-charm pentaquark \cite{Yang:2011wz} is connected with the
LHCb new measurement \cite{LHCbtalk}. For the $\Sigma_c\bar{D}$ and
$\Sigma_c\bar{D}^*$ systems, there are only three molecular states
with isospin $I=1/2$. The newly observed $P_{c}(4312)^+$ corresponds
to the $\Sigma_c\bar{D}$ with $I(J^{P})=1/2(1/2^-)$, while the
$P_c(4440)^+$ and $P_c(4457)^+$ can be identified as the
$\Sigma_c\bar{D}$ molecular states with $I(J^{P})=1/2(1/2^-)$ and
$1/2(3/2^-)$ respectively. The updated LHCb measurement
\cite{LHCbtalk} confirms the existence of molecular hidden-charm
pentaquarks composed of an S-wave charmed baryon and an S-wave
anti-charmed meson
\cite{Yang:2011wz,Wu:2010jy,Wang:2011rga,Wu:2012md,Li:2014gra,Chen:2015loa,Chen:2015moa,
Karliner:2015ina},
which shall become a milestone in the exploration of the multiquark
hadronic matter.

{\it Reproducing three pentaquarks}:---In the framework of the OBE
model, a key point is the deduction of the effective potentials for
the hadronic molecular systems. Considering the coupled channel
effect and S-D wave mixing, we may list the corresponding channel
contributions to the $\Sigma_c\bar{D}$ system with $J^P=1/2^-$, two
$\Sigma_c\bar{D}^*$ systems with $J^P=1/2^-,3/2^-$ and the
$\Sigma_c^*\bar{D}$ system with $J^P=3/2^-$, which can be
categorized into two groups
\begin{eqnarray*}
\begin{array}{cl}
1/2^-:    &\Sigma_c\bar{D}|{}^2\mathbb{S}_{\frac{1}{2}}\rangle,
    \,\Sigma_c^*\bar{D}|{}^4\mathbb{D}_{\frac{1}{2}}\rangle,
    \,\Sigma_c\bar{D}^*|{}^2\mathbb{S}_{\frac{1}{2}}/{}^4\mathbb{D}_{\frac{1}{2}}\rangle,
    \,\\&\Sigma_c^*\bar{D}^*|{}^2\mathbb{S}_{\frac{1}{2}}/
    {}^4\mathbb{D}_{\frac{1}{2}}/{}^6\mathbb{D}_{\frac{1}{2}}\rangle,\\
3/2^-:    &\Sigma_c\bar{D}^*|{}^4\mathbb{S}_{\frac{3}{2}}/
          {}^2\mathbb{D}_{\frac{3}{2}}/{}^4\mathbb{D}_{\frac{3}{2}}\rangle,
     \,\Sigma_c^*\bar{D}^*|{}^4\mathbb{S}_{\frac{3}{2}}/{}^2\mathbb{D}_{\frac{3}{2}}/
     {}^4\mathbb{D}_{\frac{3}{2}}/{}^6\mathbb{D}_{\frac{3}{2}}\rangle.
\end{array}
\end{eqnarray*}
The involved flavor wave functions $|I,I_3\rangle$ are
\begin{eqnarray}&&\begin{array}{c}
\left|\frac{1}{2},\frac{1}{2}\right\rangle =
     \sqrt{\frac{2}{3}}\left|\Sigma_c^{(*)++}{D}^{(*)-}\right\rangle
     -\frac{1}{\sqrt{3}}\left|\Sigma_c^{(*)+}\bar{D}^{(*)0}\right\rangle\\
\left|\frac{1}{2},-\frac{1}{2}\right\rangle =
     \frac{1}{\sqrt{3}}\left|\Sigma_c^{(*)+}{D}^{(*)-}\right\rangle
     -\sqrt{\frac{2}{3}}\left|\Sigma_c^{(*)0}\bar{D}^{(*)0}\right\rangle
     \end{array}.
\end{eqnarray}
In order to calculate the effective potentials of the hidden-charm
molecular baryons, we need to utilize the effective Lagrangian
describing the interactions of the charmed or anti-charmed
baryon/meson with the light mesons. The Lagrangian depicting the
interaction of the ground state charmed baryons with the light
mesons can be constructed according to the chiral symmetry, the
heavy quark symmetry and the hidden local symmetry
\cite{Liu:2011xc}, while the one for the S-wave pseudoscalar and
vector charmed mesons according to the chiral symmetry and the heavy
quark symmetry
\cite{Yan:1992gz,Wise:1992hn,Burdman:1992gh,Casalbuoni:1996pg,Falk:1992cx}.
Here we list the Lagrangians in the following
\begin{eqnarray}
\mathcal{L}_{H} &=& g_S\langle \bar{H}_a\sigma H_b\rangle
  +ig\langle \bar{H}_a\gamma_{\mu}A_{ab}^{\mu}\gamma_5H_b\rangle\nonumber\\
  &&-i\beta\langle \bar{H}_av_{\mu}\left(\mathcal{V}_{ab}^{\mu}-\rho_{ab}^{\mu}\right)
  H_b\rangle
  +i\lambda\langle \bar{H}_a\sigma_{\mu\nu}F^{\mu\nu}(\rho)H_b\rangle,\label{lag1}\\
\mathcal{L}_{\mathcal{S}}
&=&l_S\langle\bar{\mathcal{S}}_{\mu}\sigma\mathcal{S}^{\mu}\rangle
-\frac{3}{2}g_1\varepsilon^{\mu\nu\lambda\kappa}v_{\kappa}\langle\bar{\mathcal{S}}_{\mu}
A_{\nu}\mathcal{S}_{\lambda}\rangle\rangle\nonumber\\
  &&
+i\beta_{S}\langle\bar{\mathcal{S}}_{\mu}v_{\alpha}\left(\mathcal{V}_{ab}^{\alpha}
-\rho_{ab}^{\alpha}\right)
\mathcal{S}^{\mu}\rangle+\lambda_S\langle\bar{\mathcal{S}}_{\mu}F^{\mu\nu}(\rho)
\mathcal{S}_{\nu}\rangle.\label{lag2}
\end{eqnarray}
Here, the following definitions are used, i.e., $H=
[\tilde{\mathcal{P}}^{*\mu}\gamma_{\mu}-\tilde{\mathcal{P}}\gamma_5]\frac{1-\rlap\slash
v}{2}$, $\mathcal{S}_{\mu}=
-\sqrt{\frac{1}{3}}(\gamma_{\mu}+v_{\mu})\gamma^5\mathcal{B}_6
       +\mathcal{B}_{6\mu}^*$, $\bar{H}=\gamma_0H^\dagger \gamma_0$,
$\bar{\mathcal{S}}_{\mu}=\mathcal{S}_{\mu}^\dagger \gamma_0$,
$A_{\mu}=\frac{1}{2}(\xi^{\dag}\partial_{\mu}\xi-\xi\partial_{\mu}\xi^{\dag})=\frac{i}{f_{\pi}}
\partial_{\mu}\mathbb{P}+\ldots$, $\mathcal{V}_{\mu}=
\frac{1}{2}(\xi^{\dag}\partial_{\mu}\xi-\xi\partial_{\mu}\xi^{\dag})
=\frac{i}{2f_{\pi}^2}\left[\mathbb{P},\partial_{\mu}\mathbb{P}\right]+\ldots$,
$\xi=\text{exp}(i\mathbb{P}/f_{\pi})$,
$\rho^{\mu}=ig_V\mathbb{V}^{\mu}/\sqrt{2}$, and
$F^{\mu\nu}(\rho)=\partial^{\mu}\rho^{\nu}-\partial^{\nu}\rho^{\mu}
+\left[\rho^{\mu},\rho^{\nu}\right]$. And $\tilde{\mathcal{P}}$ and
$\tilde{\mathcal{P}}^*$ satisy
$\tilde{\mathcal{P}}=\left(\bar{D}^0,\,D^-\right)^T$, $\tilde{\mathcal{P}}^*=\left(\bar{D}^{*0},\,D^{*-}\right)^T$.
Note that the SU(3) chiral symmetry is considered to construct the
Lagrangians. However, for convenience, we only show the first two
lines and columns in the above matrices, since we do not consider,
in this letter, the strange case which corresponds to the third line
and column. Thus, the $\mathcal{B}_3$, $\mathcal{B}_6$ and
$\mathcal{B}_6^*$ are defined as
\begin{eqnarray*}
\mathbb{P} &=& \left(\begin{array}{cc}
\frac{\pi^0}{\sqrt{2}}+\frac{\eta}{\sqrt{6}} &\pi^+\\
\pi^- &-\frac{\pi^0}{\sqrt{2}}+\frac{\eta}{\sqrt{6}}
\end{array}\right),\quad
\mathcal{B}_6^{(*)} = \left(\begin{array}{cc}
         \Sigma_c^{{(*)}++}              &\frac{\Sigma_c^{{(*)}+}}{\sqrt{2}}\\
         \frac{\Sigma_c^{{(*)}+}}{\sqrt{2}}      &\Sigma_c^{{(*)}0}
\end{array}\right),\nonumber\\
\mathbb{V} &=& \left(\begin{array}{cc}
\frac{\rho^0}{\sqrt{2}}+\frac{\omega}{\sqrt{2}}  &\rho^+\\
\rho^- &-\frac{\rho^0}{\sqrt{2}}+\frac{\omega}{\sqrt{2}}
\end{array}\right).
\end{eqnarray*}

With the above preparation, we consider the $t$-channel Feynman
diagrams to deduce the effective potentials, where the Breit
approximation $\mathcal{V}_{E}(\bm{q}) =
          -{\mathcal{M}}/
          {\sqrt{\prod_i2M_i\prod_f2M_f}}$
is needed to relate the t-channel scattering amplitudes
$\mathcal{M}$ to the potentials $\mathcal{V}_{E}(\bm{q})$ in the
momentum space. The Fourier transformation is performed to obtain
the potentials in the coordinate space. Besides, the monopole form
factor $\mathcal{F}(q^2,m_E^2)= (\Lambda^2-m_E^2)/(\Lambda^2-q^2)$
is introduced for compensating the off shell effect of the exchanged
meson and describing the structure effect of every interaction
vertex. According to the experience of the deuteron and interaction between the proton and neutron, cutoff $\Lambda$ in the form factor is taken around 1 GeV, which is widely used to test whether a loose molecular state exists or not.

In the OBE model, The general expressions of the potentials in the coordinate
space for the discussed systems include scalar, pseudoscalar, and vector mesons exchanges part, i.e.,
\begin{eqnarray}
V^{ij}(r)=V_{\pi/\eta}^{ij}(r)+V_{\rho/\omega}^{ij}(r)+V_{\sigma}^{ij}(r),
\end{eqnarray}
according to the mass difference of exchanged mesons, $\pi$, $\sigma/\eta$, and $\rho/\omega$ exchanges contribute in long, intermediate, and short range.
For $J=1/2$, $i,j=1,2,3,4$ corresponds to the channels
$\Sigma_c\bar{D}$, \,$\Sigma_c^*\bar{D}$, \,$\Sigma_c\bar{D}^*$,
\,$\Sigma_c^*\bar{D}^*$, while for $J=3/2$, $i,j=1,2$ corresponds to
$\Sigma_c\bar{D}^*$, \,$\Sigma_c^*\bar{D}^*$. For convenience, we
define the following expressions
\begin{eqnarray}
Y(\Lambda,m,{r}) &=&\frac{1}{4\pi r}(e^{-mr}-e^{-\Lambda r})-\frac{\Lambda^2-m^2}{8\pi \Lambda}e^{-\Lambda r},\\
\mathcal{Y}^{ij}_{\Lambda m_a}&=&\mathcal{D}_{ij}Y(\Lambda,m_\sigma,r),\\
\mathcal{Z}^{ij}_{\Lambda m_a}&=&\left(\mathcal{E}_{ij}\nabla^2+\mathcal{F}_{ij}r\frac{\partial}{\partial r}\frac{1}{r}\frac{\partial}{\partial r}\right)Y(\Lambda,m_a,r),\\
\mathcal{Z}^{\prime ij}_{\Lambda
m_a}&=&\left(2\mathcal{E}_{ij}\nabla^2-\mathcal{F}_{ij}r\frac{\partial}{\partial
r}\frac{1}{r}\frac{\partial}{\partial r}\right)Y(\Lambda,m_a,r),
\end{eqnarray}
where the matrices $\mathcal{D}_{ij},\mathcal{E}_{ij}$ correspond to
the spin-spin operators and $\mathcal{F}_{ij}$ to the tensor
operator, e.g., for the process
$\Sigma_c\bar{D}^*\to\Sigma_c\bar{D}^*$,
\begin{eqnarray*}
\mathcal{D}=\bm{\epsilon}_2\cdot\bm{\epsilon}_4^{\dag},\quad
\mathcal{E}=\bm{\sigma}\cdot\left(i\bm{\epsilon}_2\times
     \bm{\epsilon}_4^{\dag}\right),\quad
\mathcal{F}=S\left(\hat{r},\bm{\sigma},
     i\bm{\epsilon}_2\times\bm{\epsilon}_4^{\dag}\right).
\end{eqnarray*}
In a numerical calculation, these operators should be sandwiched
between the discussed spin-orbit wave functions, all of the
operators will be replaced by serials of nonzero matrix elements,
for example, for $\Sigma_c\bar{D}^*\to\Sigma_c\bar{D}^*$ with
$J^P=1/2^-$, $\langle
\Sigma_c\bar{D}^*(1/2^-)|\mathcal{E}|\Sigma_c\bar{D}^*(1/2^-)\rangle
=\left(\begin{array}{cc}-2 &0\\0 &1\end{array}\right)$. The
subpotentials now can be written as
\begin{eqnarray}
V^{11}&=&-AY(\Lambda,m_{\sigma},r)
       -\frac{\mathcal{G}B}{2}Y(\Lambda,m_{\rho},r)-\frac{B}{4}Y(\Lambda,m_{\omega},r),\nonumber\\
V^{12}&=&\frac{A}{\sqrt{3}}\mathcal{Y}_{\Lambda_3 m_{\sigma3}}
       +\frac{\mathcal{G}B}{2\sqrt{3}}\mathcal{Y}^{12}_{\Lambda_3,m_{\rho3}}
       +\frac{B}{4\sqrt{3}}\mathcal{Y}^{12}_{\Lambda_3,m_{\omega3}},\nonumber\\
V^{13} &=&
        \frac{\mathcal{G}C}{3}\mathcal{Z}^{13}_{\Lambda_4,m_{\pi4}}
        +\frac{C}{18}\mathcal{Z}^{13}_{\Lambda_4,m_{\eta4}}
        +\frac{2\mathcal{G}D}{9}\mathcal{Z}^{\prime13}_{\Lambda_4,m_{\rho4}}\nonumber\\&&
        +\frac{D}{9}\mathcal{Z}^{\prime13}_{\Lambda_4,m_{\omega4}},\nonumber\\
V^{14}
&=&\frac{\mathcal{G}C}{2\sqrt{3}}\mathcal{Z}^{14}_{\Lambda_5,m_{\pi5}}
        +\frac{C}{12\sqrt{3}}\mathcal{Z}^{14}_{\Lambda_5,m_{\eta5}}
        +\frac{\mathcal{G}D}{3\sqrt{3}}\mathcal{Z}^{\prime14}_{\Lambda_5,m_{\rho5}}\nonumber\\
        &&
        +\frac{D}{6\sqrt{3}}\mathcal{Z}^{\prime14}_{\Lambda_5,m_{\omega5}},\nonumber\\
V^{22} &=&-A\mathcal{Y}^{11}_{\Lambda,m_{\sigma}}
       -\frac{\mathcal{G}B}{2}\mathcal{Y}^{22}_{\Lambda,m_{\rho}}
       -\frac{B}{4}\mathcal{Y}^{22}_{\Lambda,m_{\omega}},\nonumber\\
V^{23}&=&\frac{\mathcal{G}C}{2\sqrt{3}}
\mathcal{Z}^{23}_{\Lambda_0,m_{\pi0}}
    +\frac{C}{12\sqrt{3}}\mathcal{Z}^{23}_{\Lambda_0,m_{\eta0}}
    -\frac{\mathcal{G}D}{3\sqrt{3}}\mathcal{Z}^{\prime23}_{\Lambda_0,m_{\rho0}}\nonumber\\
    &&
    -\frac{D}{6\sqrt{3}}\mathcal{Z}^{\prime23}_{\Lambda_0,m_{\omega0}},\nonumber\\
V^{24}&=&
    \frac{\mathcal{G}C}{2}\mathcal{Z}^{24}_{\Lambda_1,m_{\pi1}}
    +\frac{C}{12}\mathcal{Z}^{24}_{\Lambda_1,m_{\eta1}}
    -\frac{\mathcal{G}D}{3}\mathcal{Z}^{\prime24}_{\Lambda_1,m_{\rho1}}\nonumber\\
    &&
    -\frac{D}{6}\mathcal{Z}^{\prime 24}_{\Lambda_1,m_{\omega1}},\nonumber\\
V^{33}&=&-A\mathcal{Y}^{33}_{\Lambda,m_{\sigma}}
     +\frac{\mathcal{G}C}{3}\mathcal{Z}^{33}_{\Lambda,m_{\pi}}
     +\frac{C}{18}\mathcal{Z}^{33}_{\Lambda,m_{\eta}}
    -\frac{\mathcal{G}B}{2}\mathcal{Y}^{33}_{\Lambda,m_{\rho}}\nonumber\\
    &&-\frac{2\mathcal{G}D}{9}\mathcal{Z}^{\prime33}_{\Lambda,m_{\rho}}
    -\frac{B}{4}\mathcal{Y}^{33}_{\Lambda,m_{\omega}}
    -\frac{D}{9}\mathcal{Z}^{\prime33}_{\Lambda,m_{\omega}},\nonumber\\
V^{34}&=&\frac{A}{\sqrt{3}}\mathcal{Y}^{34}_{\Lambda_2,m_{\sigma2}}
    +\frac{\sqrt{3}\mathcal{G}C}{6}\mathcal{Z}^{34}_{\Lambda_2,m_{\pi2}}
    +\frac{\sqrt{3}C}{36}\mathcal{Z}^{34}_{\Lambda_2,m_{\eta2}}\nonumber\\
        &&
        +\frac{\mathcal{G}B}{2\sqrt{3}}\mathcal{Y}^{34}_{\Lambda_2,m_{\rho2}}
        -\frac{\mathcal{G}D}{3\sqrt{3}}\mathcal{Z}^{\prime34}_{\Lambda_2,m_{\rho2}}
        +\frac{B}{4\sqrt{3}}\mathcal{Y}^{34}_{\Lambda_2,m_{\omega2}}\nonumber\\
        &&-\frac{D}{6\sqrt{3}}\mathcal{Z}^{\prime 34}_{\Lambda_2,m_{\omega2}},\nonumber\\
V^{44}&=& -A\mathcal{Y}^{44}_{\Lambda,m_{\sigma}}
       -\frac{\mathcal{G}C}{2}\mathcal{Z}^{44}_{\Lambda,m_{\pi}}
       -\frac{C}{12}\mathcal{Z}^{44}_{\Lambda,m_{\eta}}
       -\frac{\mathcal{G}B}{2}\mathcal{Y}^{44}_{\Lambda,m_{\rho}}\nonumber\\
       &&+\frac{\mathcal{G}D}{3}\mathcal{Z}^{\prime44}_{\Lambda,m_{\rho}}¡¢
       -\frac{B}{4}\mathcal{Y}^{44}_{\Lambda,m_{\omega}}
       +\frac{D}{6}\mathcal{Z}^{\prime44}_{\Lambda,m_{\omega}}.
\end{eqnarray}
Here, $A=l_Sg_S$, $B=\beta\beta_Sg_V^2$, $C=\frac{gg_1}{f_\pi^2}$,
$D=\lambda\lambda_Sg_V^2$, $\mathcal{G}$ is the isospin factor,
which is taken as $-1$ for the isospin $1/2$ system (for the isospin
$3/2$ system $\mathcal{G}=1/2$). The variables in these functions
are defined as $\Lambda_b^2 =\Lambda^2-q_b^2$,
$m_{{b}}^2=m^2-q_b^2$, with $b=0,1,...,5$, and $q_i^2$s are the
functions of the masses of the charmed mesons and baryons whose
values are $q_0=35.60$ MeV, $q_1=61.41$ MeV, $q_2=25.08$ MeV,
$q_3=97.38$ MeV, $q_4=60.53$ MeV, $q_5=35.45$ MeV.

The values of the coupling constants are taken from
\cite{Liu:2007bf,Liu:2011xc,Casalbuoni:1996pg,Falk:1992cx},
$g_S=0.76,g=0.59,\beta=0.9,l_S=6.2,g_1=0.94,\beta_S=-1.74,$
$\lambda=0.56$ GeV$^{-1}$, $\lambda_S=-3.31$ GeV$^{-1}$, and $g_V=5.9$. Here, the coupling constants for the pion exchange $g$ and $g_1$ are determined from the experiment data, $\Gamma(D^*\to D\pi)$ and $\Gamma(\Sigma_c^{(*)}\to \Lambda_c\pi)$ \cite{Tanabashi:2018oca}. For the scalar and vector mesons exchanges, the coupling constants are estimated in the quark model, where the average of the coupling vertex is simultaneously calculated, at hadron level and at quark level, and the heavy quark does not couple to the light mesons. The signs are fixed with the help of the quark model.

\begin{figure}[!htbp]
\center
\includegraphics[width=3.3in]{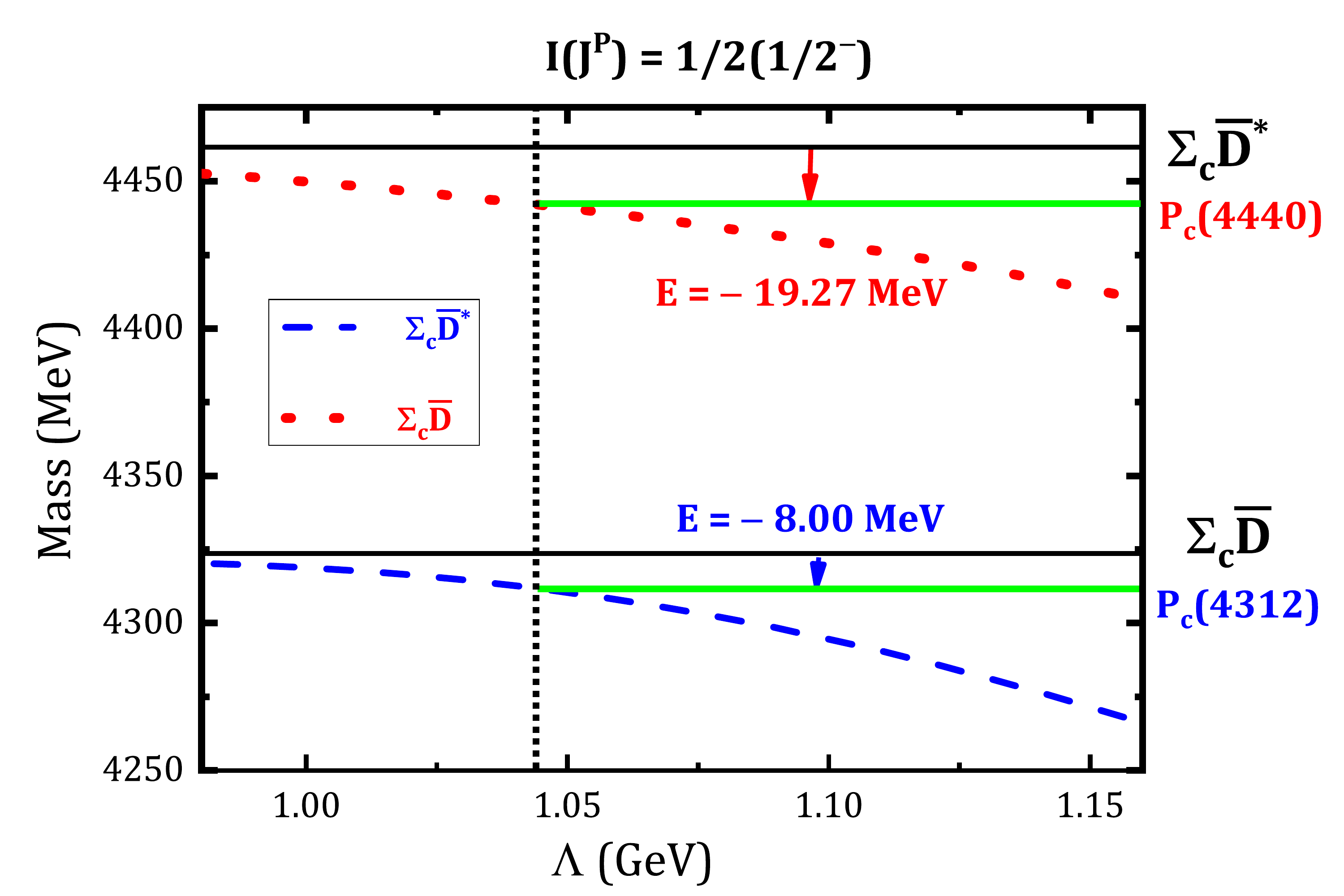}
\includegraphics[width=3.3in]{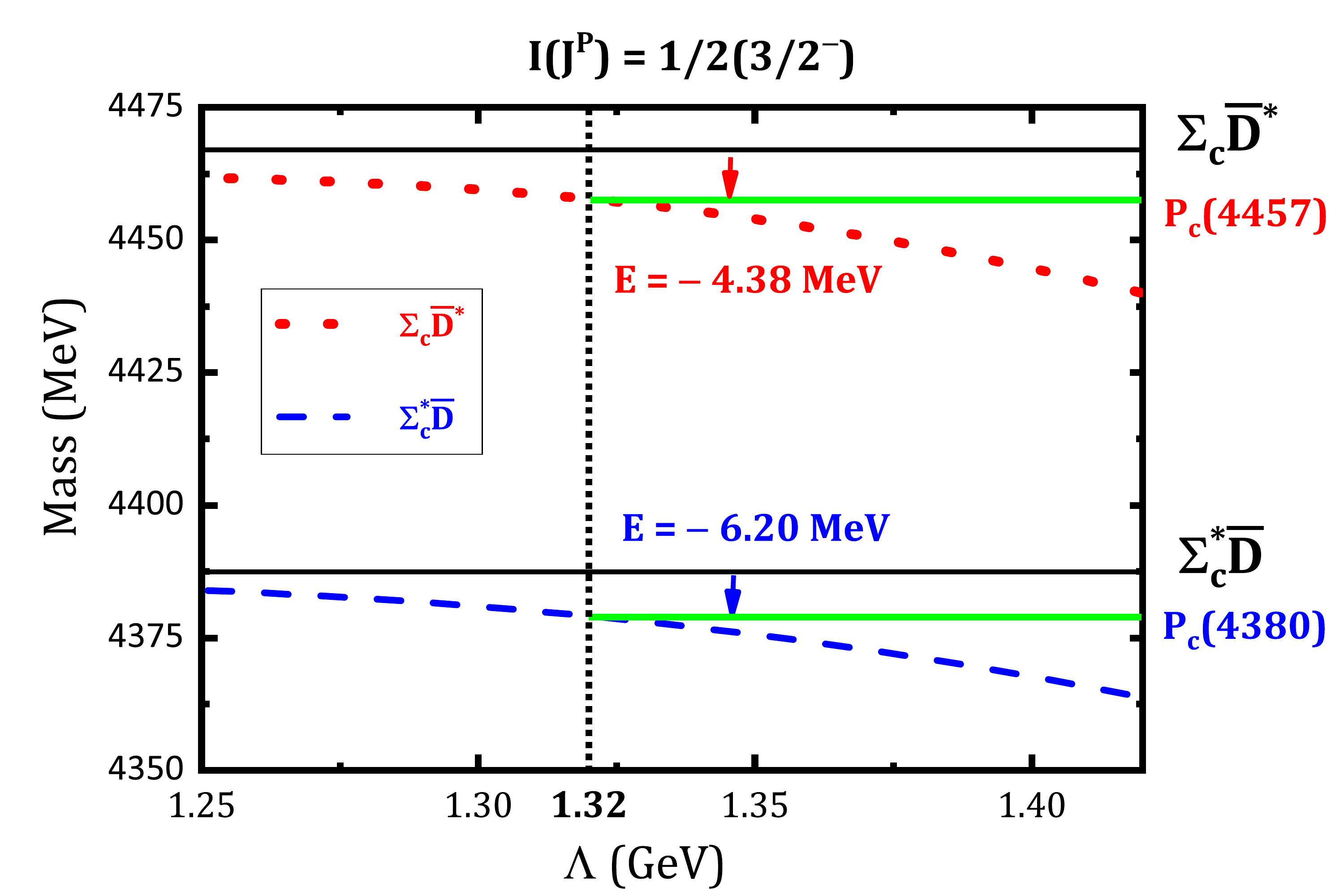}
\caption{The $\Lambda$ dependence of the bound state masses for the
coupled channel $\Sigma_c^{(*)}\bar{D}^{(*)}$ systems with $I=1/2$.
Here, $E$ is the binding energy. The vertical dotted lines and
horizontal solid lines stand for the cutoff $\Lambda$ value and
hidden-charm meson-baryon systems thresholds.}\label{mass}
\end{figure}

Finally, the binding energies for the $\Sigma_c\bar{D}$ and
$\Sigma_c\bar{D}^*$ systems can be evaluated by solving the
Schr$\rm\ddot{o}$dinger equation. In Fig. \ref{mass}, we present
{the bound state masses for the coupled channel
$\Sigma_c^{(*)}\bar{D}^{(*)}$ systems with $I=1/2$ which depends on
the cutoff $\Lambda$. As the cutoff increases, the binding energy of
all the systems becomes deeper and deeper.}

For the $J^P=1/2^-$ case, we can reproduce the masses of the
$P_c(4312)$ and $P_c(4440)$ with a very reasonable cutoff value
$\Lambda=1.04$ GeV. {The} binding energies for these systems are
$-8.00$ MeV and $-19.27$ MeV, respectively.
Their corresponding root-mean-square (RMS) radii are 1.22 fm and 0.88 fm, respectively.
For the $P_c(4312)$, the
$\Sigma_c\bar{D}$ channel is dominant, the probabilities of the
$\Sigma_c\bar{D}$, $\Sigma_c\bar{D}^*$, and $\Sigma_c^*\bar{D}^*$
components are around 84\%, 12\%, and 4\%, respectively. The
$P_c(4440)$ is mainly composed of the S-wave $\Sigma_c\bar{D}^*$
component with the probability over 94\%.

{For the $I(J^P)=1/2(3/2^-)$ case, as shown in Fig. \ref{mass},}
the mass of the $P_c(4457)$ is also reproduced with $\Lambda=1.32$
GeV. Here, its binding energy and RMS radius are $-4.38$ MeV and 1.61 fm, respectively. The coupled-channel effect plays an important role,
{since} the probabilities of the $\Sigma_c\bar{D}^*$ and
$\Sigma_c^*\bar{D}^*$ {components} are around 75\% and 25\%,
respectively.

With the same {value of the cutoff as that of $P_c(4457)$}, we find
{another} bound state {of the quantum number $I(J^P)=1/2(3/2^-)$,
which locates below the $\Sigma_c^*\bar{D}$ threshold with a binding
energy $E=-6.20$ MeV and RMS radius $r_{RMS}=1.40$ fm. The dominant channel is the
$\Sigma_c^*\bar{D}({}^4S_{3/2})$ with a probability over 87\%. This
state can be identified with the $P_c^+(4380)$ observed in 2015 also
by LHCb collaboration \cite{Aaij:2015tga}, although the recent LHCb
measurement doesn't give any information of such a particle. We hope
the further experiments can provide a more precise measurement and
confirm the existence of this state.}

{From the above discussion, we see that the three resonances
$P_c^+(4312)$, $P_c^+(4440)$, $P_c^+(4457)$ are naturally
interpreted within the loosely bound $\Sigma_c\bar{D}$ and
$\Sigma_c\bar{D}^*$ molecule picture. For the $\Sigma_c
\bar{D}^{(*)}$ systems, there are only three molecular states with
isospin $1/2$, i.e., $\Sigma_c \bar{D}$ corresponding to the lower
peak $P_c(4312)^+$, and $\Sigma_c \bar{D}^*$ to the higher peaks
$P_c(4440)$ and $P_c(4457)$. Within the hidden-charm molecular
assignment, we can easily understand why the two higher peaks are
very close to each other, and the mass difference between the two
higher peaks and the lowest $P_c^+(4312)$ peak is around $140$ MeV
which is approximately the mass difference of the $D$ and $D^*$
mesons.}


{\it More predictions}:--- Besides explaining the three
{particles $P_c(4312)$, $P_c(4440)$ and $P_c(4457)$, we
further investigate the possible existence of the
$\Sigma_c^{(*)}\bar{D}^{(*)}$ with
$(I=\frac{3}{2},J^P=\frac{1}{2}^-/\frac{3}{2}^-)$. In Table
\ref{num}, we collect {the corresponding bound state
solutions.} However, when the bound state solutions emerge, the
corresponding cutoffs are around $2\sim 4$ GeV and quite large for
all these systems. The past experience indicates that such a large
cutoff value sometimes points to the non-existence of the molecular
states. Consequently, we cannot give a clear conclusion whether the
$\Sigma_c^{(*)}\bar{D}^{(*)}$ systems with isospin $3/2$ exist or
not.}

\renewcommand\tabcolsep{0.35cm}
\renewcommand{\arraystretch}{1.8}
\begin{table}[!htbp]
  \centering
  \caption{Bound state solutions for the possible hidden-charm molecular pentaquarks $\Sigma_c^{(*)}\bar{D}^{(*)}$ with $I=3/2$. The units for cutoff $\Lambda$, $E$, and root-mean-square radius $r_{RMS}$ are GeV, MeV, and fm, respectively.}\label{num}
  \begin{tabular}{ccc|ccc}
    \toprule[1pt]
  $\Lambda$    &$E$    &$r_{RMS}$     &$\Lambda$    &$E$    &$r_{RMS}$  \\\hline
  \multicolumn{3}{c|}{$\Sigma_c\bar{D}\, J^P=1/2^-$}
  &\multicolumn{3}{c}{$\Sigma_c^*\bar{D}\, J^P=3/2^-$}\\
  2.26     &$-0.22$      &4.98     &3.04     &$-1.85$      &0.80\\
  2.32     &$-5.40$      &1.35     &3.05     &$-6.03$      &0.49\\
  2.38     &$-16.17$     &0.82     &3.06     &$-10.43$     &0.42\\\hline
  \multicolumn{3}{c|}{$\Sigma_c\bar{D}^*\, J^P=1/2^-$}
  &\multicolumn{3}{c}{$\Sigma_c\bar{D}^*\, J^P=3/2^-$}\\
  4.21     &$-0.63$      &2.57     &2.97    &$-0.41$      &4.18\\
  4.23     &$-5.48$      &0.77     &3.03    &$-4.80$      &1.38\\
  4.25     &$-11.33$     &0.58     &3.09    &$-13.46$     &0.86\\\bottomrule[1pt]
  \end{tabular}
\end{table}

{In the experimental process $\Lambda_b\to J/\psi p K$,} all the
$P_c$ states are produced by the weak decay of $\Lambda_b^0$. With
the assignment that the $P_c(4312)$ and $P_c(4440)/P_c(4457)$ are
the $\Sigma_c\bar{D}$ and $\Sigma_c\bar{D}^*$ molecules,
respectively, these states can transit to the final $J/\psi p$ by
exchanging a S-wave charmed meson{, and the $J/\psi$ and $p$
reside in a S wave.}

The molecular systems $\Sigma_c^{(*)}\bar{D}^{(*)}$ with{
$(I=\frac{3}{2},J^P=\frac{1}{2}^-/\frac{3}{2}^-)$, if they should
exist, can not decay into $J/\psi p$ due to {the isospin}
conservation. A possible decay mode is $J/\psi \Delta(1232)$ for
both $J=1/2$ and $3/2$, which will be helpful for future
experimental search.}

{\it Summary}:---Very recently, LHCb reported the observation of
three new pentaquarks $P_c(4312)^+$, $P_c(4440)^+$ and $P_c(4457)^+$
in the $J/\psi p$ invariant mass spectrum of $\Lambda_b\to J/\psi p
K$ \cite{LHCbtalk}, which benefitted from large data samples
collected at LHC Run 1 and Run 2. This new discovery further opens
the mysterious Pandora's box of exploring exotic multiquark states.
This new LHCb observation the first time confirmed the existence of
the baryon-meson configuration molecular hidden-charm pentaquarks in
experiment, which should become a milestone of the exploration of
the multiquark hadronic matter. In this work, we have performed a
phenomenological study to decode the inner structure of the newly
observed $P_c(4312)$, $P_c(4440)$ and $P_c(4457)$ based on the OBE
model after the inclusion of the coupled-channel effect and S-D wave
mixing. The consistence of the LHCb new observation with the
hidden-charm molecular pentaquarks with the baryon-meson
configuration is clearly demonstrated.

In this work, we have also studied the interaction of the
$\Sigma^*_c\bar{D}$ system in the framework of the OBE model, and
found the existence of a $\Sigma^*_c\bar{D}$ molecular hidden-charm
pentaquark with $I(J^P)=1/2(3/2^-)$, which has a mass 4379.11 MeV.
This predicted $\Sigma^*_c\bar{D}$ molecular hidden-charm pentaquark
may correspond to the $P_c(4380)$ reported by LHCb in 2015
\cite{Aaij:2015tga}. However, the talk at the Rencontres de Moriond
QCD conference \cite{LHCbtalk} did not mention $P_c(4380)$. After
checking the new experimental data released by LHCb \cite{Aaij:2019vzc},
one may notice the evidence of the accumulation of events around
4380 MeV in the $J/\psi p$ invariant mass spectrum. Moreover, we
have also studied the possible existence of the hidden-charm
molecular pentaquarks with isospin $I=3/2$, which are the isospin
partners of the newly observed $P_c(4312)^+$, $P_c(4440)^+$ and
$P_c(4457)^+$. These states should also be searched for at LHCb. The
past sixteen years has been the golden age of the search of the
exotic hadrons. 

\section*{ACKNOWLEDGMENTS}

This project is supported by the National Natural Science Foundation
of China under Grants No. 11705069 and No. 11575008, and National Key
Basic Research Program of China (Grant No. 2015CB856700). R.C. is also supported by the National Postdoctoral Program for Innovative Talent. X.L. is also supported
by the National Program for Support of Top-notch Young Professionals
and the China National Funds for Distinguished Young Scientists
under Grant No. 11825503.

\end{document}